\documentclass[aps,twocolumn,showpacs,amsmath,amssymb,pra,superscriptaddress,floatfix,longbibliography]{revtex4-1}

\usepackage{graphicx}    
\usepackage{hyperref} 
\usepackage{xcolor} 
\usepackage{physics} 
\usepackage{subcaption} 
\usepackage{ragged2e}
\usepackage[justification=justified, singlelinecheck=false]{caption}
\usepackage{bm}
\usepackage{booktabs}
\usepackage{soul}

\usepackage{xargs}
\usepackage[colorinlistoftodos,prependcaption,textsize=tiny]{todonotes}
\newcommandx{\tohfa}[2][1=]{\todo[linecolor=OrangeRed,backgroundcolor=OrangeRed!25,bordercolor=red,#1]{#2}}
\newcommandx{\fr}[2][1=]{\todo[linecolor=Blue,backgroundcolor=Blue!25,bordercolor=blue,#1]{#2}}

\newcommand{\imu}{\text{\rm i}}
\newcommand{\diff}{\text{\rm d}}

\newcommand{\spindown}[1]{\hat\sigma_{#1}}
\newcommand{\spinup}[1]{\spindown{#1}^\dagger}
\newcommand{\exc}[1]{\hat e_{#1}}

\newcommand{\fieldmode}[2]{\hat{#1}_{#2}}

\newcommand{\fieldmodeatom}[1]{\fieldmode{b}{#1}}
\newcommand{\inmodeatom}[1]{\fieldmodeatom{#1}^\text{in}}

\newcommand{\wgmode}[1]{\fieldmode{a}{#1}}
\newcommand{\wgmodedag}[1]{\fieldmode{a}{#1}^\dagger}
\newcommand{\wgbmode}[1]{\fieldmode{b}{#1}}
\newcommand{\wgbmodedag}[1]{\fieldmode{b}{#1}^\dagger}

\newcommand{\betab}{\beta^-}
\newcommand{\betaf}{\beta^+}

\newcommand{\betaUni}{\beta}

\newcommand{\density}{\hat {\rho}}

\newcommand{\ddt}{\frac{d}{dt}}
\newcommand{\ha}{\text{h.c.}}

\renewcommand{\vec}[1]{\boldsymbol{#1}}

\newcommand{\nth}[1]{#1^\text{th}}
\newcommand{\expect}[1]{\langle #1 \rangle}
\newcommand{\Eq}[1]{Eq.~(#1)}
\newcommand{\Rho}{\mathfrak{P}}
\newcommand{\allexcited}{\ket{ee\dots e}}

\begin{document}
\title {Photon statistics in chiral waveguide QED: I Mean field and perturbative expansions}

\author{M. Eltohfa}
 \email{meltohfa@purdue.edu}
 \affiliation{
 Department of Physics and Astronomy, Purdue University, West Lafayette, Indiana 47906 USA
}
\author{F. Robicheaux}%
 \email{robichf@purdue.edu}
\affiliation{
 Department of Physics and Astronomy, Purdue University, West Lafayette, Indiana 47906 USA
}
\affiliation{Purdue Quantum Science and Engineering Institute, Purdue
University, West Lafayette, Indiana 47907, USA}
\date{\today}

\begin{abstract}
    Waveguide Quantum Electrodynamics (WQED) offers a suitable stage for controlling the interaction of light with atoms, allowing for collective phenomena such as super- and subradiance. In a chiral waveguide setup, the quantum state evolves through all the Hilbert space, rendering an exact theoretical treatment exponentially hard and unobtained to date for more than $\sim 20$ atoms. In this work, we use a computationally efficient higher order mean-field approximation to model the radiation dynamics in a chirally coupled array of atoms, showing good agreement with recent experimental results. Further, based on a perturbative approximation of the full dynamics, we develop an analytical solution that captures photon statistics for a moderate atom number, $N$, and a homogeneous atom-waveguide coupling, $\beta$. Finally, we show that capturing the onset of second-order coherence from a fully inverted state requires a fourth-order mean-field approximation, as lower-order treatments fail to account for the necessary four-body correlations. These results illustrate the complex behavior of a symmetry-lacking system, and the methods discussed here provide systematic analytical solutions to which semi-classical methods such as the cumulant expansion or the truncated Wigner approximation can be benchmarked.
\end{abstract}


\maketitle

\section{Introduction}\label{sec:intro}
The study of many-body quantum systems has been a crucial task because of their rich physical phenomena. In the context of quantum optics, light interacts with an ensemble of quantum emitters leading to various effects, including super- and subradiance \cite{dicke1954coherence,gross1982superradiance,gross1976observation,scully2009collective,pellegrino2014observation,prasad2000polarium,devoe1996observation} and the manipulation of atomic or light states \cite{guimond2019subradiant,jen2024photon, hammerer2010quantum, PhysRevLett.119.053901, PhysRevLett.117.243601, PhysRevA.97.023833, bettles2020quantum} that is useful for coherent quantum control and quantum information processing. However, many-body quantum systems can be intrinsically complex due to the exponentially growing number of states.

In this work, we aim to tackle this complexity in a many-atom system within the framework of waveguide quantum electrodynamics (WQED). WQED offers a suitable environment to achieve the aforementioned phenomena which have been demonstrated in various theoretical \cite{asenjo2017exponential,TWA2024cascaded,anaPhysRevLett.131.033605,Retardation2025effects,MFchiral2023higherorder} and experimental \cite{setupPhysRevLett.104.203603,PRLLiedl2023collective,PRXQuantum.4.030304,bach2024emergence} studies. In particular, we focus on the statistics of the photons emitted from an inverted ensemble. This includes the photon flux and two-time photon-photon correlation functions during the superradiant burst, which have been experimentally demonstrated \cite{bach2024emergence}. Unlike in the Dicke limit of a closely packed ensemble, where there is a permutational symmetry between all atoms, a general WQED setup lacks any symmetry \cite{MFchiral2023higherorder} resulting in the system exploring all the available states of the Hilbert space. This was demonstrated in the single-excitation manifold \cite{oscillationPhysRevLett.128.203601}. In the many-excitation scenario considered in this work, this implies the need to simulate an exponentially growing Hilbert space to obtain exact solutions. This works only for small atom number, $N \lesssim 20$, so we turn to the use of various approximations to simulate experimentally accessible atom numbers, $ N \sim 10^3$.

In the context of WQED, various theoretical approaches have been employed. Features of interest such as the superradiant burst, emergence of second order coherence, and atomic correlations can be extracted with tools such as the cumulant expansion (also known as higher order mean field or MF) methods \cite{MFchiral2023higherorder}, the truncated Wigner approximation (TWA) \cite{TWA2024cascaded, Retardation2025effects}, matrix product states (MPS) \cite{Retardation2025effects}, and the mixed coherent state approximation (MCSA) \cite{PRXsuperradiant2024bursts}. These tools vary not only in their complexity but also in what quantities they are able to capture reliably. So far, the \emph{two-time} photon correlations, which have been experimentally accessible \cite{bach2024emergence}, have not been demonstrated using these techniques.

In this work, we are motivated to model the recent experimental results in Ref.~\cite{bach2024emergence} regarding the \emph{two-time} photon correlations. We develop higher order MF methods that are computationally efficient owing to the one-dimensional (1D) nature of the system. This allows us to explore systems with many atoms ($N$ reaching tens of thousands as done in a companion paper \cite{tohfathermlimit}) and generate results that qualitatively agree with the experiment.

To establish the validity of the MF approach, we compare our method to exact calculations based on density matrix (DM) evolution or Monte Carlo (MC) trajectories for small atom numbers ($\lessapprox 20$), showing good agreement. Additionally, we utilize the small atom-waveguide coupling, $\beta$, ($\approx 0.01$ in the experiment in \cite{bach2024emergence}) to extract an analytical solution via a Neumann expansion. This expansion sheds light on the multiple timescales present as the state of the system navigates all the states of the Hilbert space. It also allows us to benchmark semi-classical methods such as the MF or the TWA approximation at moderate atom numbers, $N \sim 10^3$. In a companion paper \cite{tohfathermlimit}, we explore the thermodynamic limit as $N \rightarrow \infty$ keeping the optical depth, $OD = 4 N \beta$, fixed. There, the multiple timescales collapse revealing a closed form for various observables and predicting the absence of second-order coherence in such limit.

We note that for the experimental modeling, our methods take into account the \emph{fluctuation} of the atom-waveguide coupling, $\beta$, from one atom to another. On the other hand, for calculations done to compare the behavior and scaling of the different methods, we assume the \emph{same} $\beta$ for each atom.



This paper is organized as follows. In Sec.~\ref{sec:methods}, we define the \emph{chiral} system under investigation.
In Sec.~\ref{sec:mf_equation}, we introduce the mean field approach for WQED and a way to make it computationally efficient. In Sec.~\ref{sec:analytical_methods}, we introduce an analytical solution based on a perturbative expansion of the exact solution starting from a symmetric initial condition. In Sec.~\ref{sec:results}, we benchmark the MF and the analytic methods. In Sec.~\ref{sec:simulation_experiment}, we present some experimental considerations and compare experimental results to predictions by a second order mean field (MF2) method. Finally, in Sec.~\ref{sec:coherence_full_inverted} we present a shortcoming of MF2 in the case of starting from idealized inversion and its possible impact on successfully modeling the experiment.



\section{Methods}
\label{sec:methods}

\begin{figure}[htbp]
    \centering
    \includegraphics[width=0.45\textwidth]{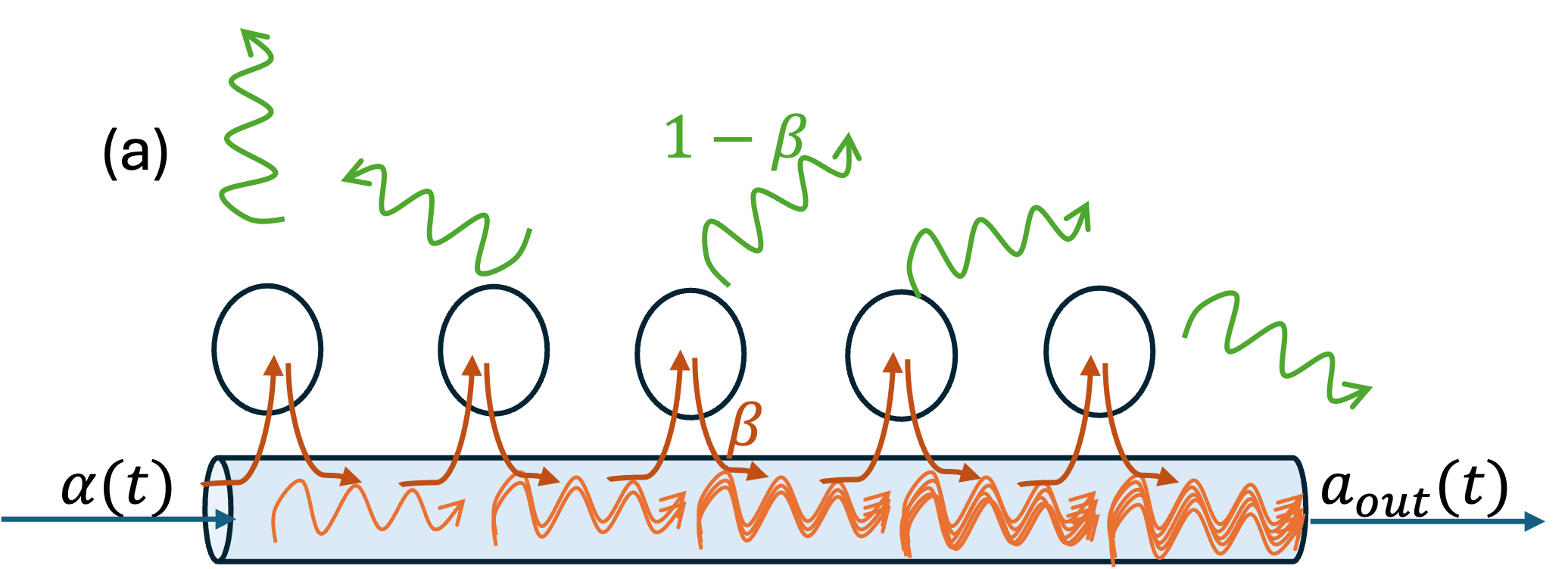}
    \includegraphics[width=0.45\textwidth]{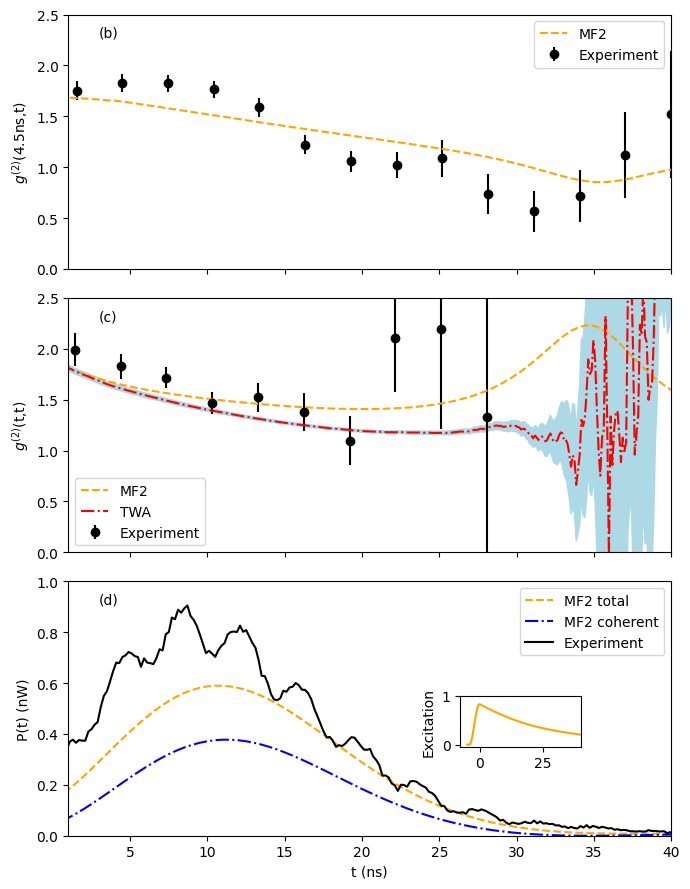}
    \caption{\justifying
        (a) Schematic of the setup. $N$ two-level atoms are coupled to a unidirectional waveguide. 
        The coupling constant to the right waveguided mode is $\beta$ and to freespace modes is $1 - \beta$. This realizes a cascaded quantum system.
        We are interested in correlators of the output mode $\wgmode{\text{out}}$, such as the output power $P(t) = \expect{\wgmode{\text{out}}^\dagger(t) \wgmode{\text{out}}(t)}$ and Glauber's second-order quantum correlation function $G^{(2)}(t_1,t_2) = \expect{\wgmode{\text{out}}^\dagger(t_1) \wgmode{\text{out}}^\dagger(t_2)  \wgmode{\text{out}}(t_2)\wgmode{\text{out}}(t_1)}$.
        In Ref.~\cite{bach2024emergence}, this model is implemented using the D2 transition of nanofiber-coupled cold cesium atoms with excited state lifetime 30.5~ns (b) We show experimental data from Ref.~\cite{bach2024emergence} $g^{(2)}(4.5 \,\text{ns},t) = G^{(2)}(4.5\, \text{ns},t) / (P(4.5\, \text{ns})P(t))$ (black data points) for the ensemble decaying from near maximal inversion alongside the theoretical prediction by a second order cumulant expansion or mean field (MF2) method as laid out in this work. (c) we show experimental data for $g^{(2)}(t,t)$ as well as predictions by a MF2 method and the truncated Wigner approximation (TWA) used in Refs.~\cite{TWA2024cascaded,bach2024emergence}. The shaded lightblue area indicates the statistical uncertainty of the TWA method.
        (d) We show experimental data of $P(t)$ alongside the mean field prediction. We also show the theoretically predicted coherent component of $P(t)$, as explained in the text. The inset shows the excited population fraction during and after the inversion pulse as predicted by MF2. Details of the experimental parameters are given in Sec.~\ref{sec:simulation_experiment}.
        }
    \label{fig:intro_figure}
\end{figure}

The system under consideration is described in detail in Refs.~\cite{PRLLiedl2023collective,PRXsuperradiant2024bursts,TWA2024cascaded} and is shown in the top panel of Fig.~\ref{fig:intro_figure}. It consists of a waveguide with $N$ atoms trapped in its vicinity. We denote the ordered atom positions along the waveguide as $z_i$ for $ 1 \leq i \leq N$ such that $z_i < z_j$ if $i < j$. A single atom radiates spontaneously with rate $\Gamma_0$ which branches into the waveguide's forward and backward directions and into freespace modes with probabilities $\betaf$, $\betab$, and $1 - \betaf - \betab$ respectively. After applying a magnetic field, and using circularly polarized light, the atoms can be made to couple mainly to the right-propagating photons of the waveguide, i.e., $\betaf \gg \betab$ \cite{PRXsuperradiant2024bursts}. In the rest of the main text, we assume a unidirectional waveguide ($\betab = 0$), and denote $\betaf \equiv \betaUni$, although some of the methods we use are easily extended to a bidirectional waveguide as shown in App.~\ref{sec:bidirectional}.


To study the collective spontaneous radiation, the atoms are driven by a short resonant laser pulse in the waveguide that addresses only two internal states, which are denoted by $\ket{g}$ and $\ket{e}$ for the ground and excited states respectively. After the pulse, the atoms are partially or maximally inverted and then are let to decay spontaneously. The photons emitted in the waveguide are captured by detectors at the right end, using which the photon intensity or intensity-intensity correlations can be measured.


As is commonly considered, we work in the Markovian picture where the quantized light field is traced out. In addition, a rotating frame with the transition frequency ($\omega_0$) is chosen. As explained in Ref.~\cite{TWA2024cascaded}, the locations of the atoms can also be absorbed by another rotating frame in the case of a unidirectional waveguide. The resulting Lindblad master equation for the density matrix $\density$ reads \cite{TWA2024cascaded}
\begin{equation}\label{eq:master}
    \ddt \density = -\imu [\hat H_0 + \hat H_\text{casc}, \density] + \mathcal{L}_\text{coll}[\density] + \mathcal{L}_0[\density]
\end{equation}
with ($\ha$ stands for Hermitian conjugate)
\begin{subequations} \label{eq:cascaded_superoperators}
\begin{align}
    \hat H_0 &=  \sum_n  \sqrt{\beta_n} \left( \alpha(t) \spinup{n} + \ha \right) ,\\
    \hat H_\text{casc} &= -\frac{\imu}{2} \sum_m \sum_{n<m} \left( \sqrt{\beta_m \beta_n} \spinup{m}\spindown{n} - \ha \right), \label{eq:cascadedHamiltonian}\\
    \mathcal{L}_\text{coll}[\density] &= \sum_{m,n} \sqrt{\beta_m \beta_n} \left( \spindown{n} \density \spinup{m} - \frac{1}{2} \left\{ \spinup{m}\spindown{n}, \density \right\} \right),  \label{eq:collectivedecay}\\
    \mathcal{L}_0[\density] &=  \sum_n \left(1 - \beta_n \right) \left( \spindown{n} \density \spinup{n} - \frac{1}{2} \left\{ \spinup{n}\spindown{n}, \density \right\} \right), 
\end{align}
\end{subequations} where we set $\hbar = 1$ and work in units of the lifetime of the excited state such that $\Gamma_0 = 1$. In Eqs.~(\ref{eq:cascaded_superoperators}), $\hat H_0$ describes the coherent resonant drive, where $\alpha(t)$ is proportional to the field amplitude of the input laser (see Fig.~\ref{fig:intro_figure}(a)). The input pulse power can then be written as $P_p(t) = \hbar \omega_0 \Gamma_0 |\alpha(t)|^2$, where $\omega_0$ is the resonance frequency. Further, $\beta_n$ denotes the coupling of the $\nth{n}$ atom to waveguide, which generally varies from one atom to another. Further, $\hat H_\text{casc}$ and $\mathcal{L}_\text{coll}$ describe the cascaded coupling of the atoms to the waveguide and are responsible for collective dynamics and correlations among atoms. $\mathcal{L}_0$ describes the individual decay of atoms to freespace. $\spindown{n} = \ket{g_n} \bra{e_n}$ is the spin-lowering operator of the $\nth{n}$ atom. Furthermore, the excitation operator of the $\nth{n}$ atom is given by $\hat{e}_n = \spinup{n}\spindown{n}$. 


It is vital to consider a fluctuation of $\beta_i$ from an atom to another when modeling the experiment (see Refs.\cite{TWA2024cascaded,bach2024emergence}). A varying $\beta$ was employed for the results in Fig.~\ref{fig:intro_figure} and in Sec.~\ref{sec:simulation_experiment}, but is ignored throughout calculations that compare and study the trends in different approaches.




The goal of this work is to find numerical predictions for time-dependent correlators of the output of the waveguide,
\begin{equation}\label{eq:output}
    \wgmode{\text{out}} = \alpha - \imu \sum_n \sqrt{\beta_n} \spindown{n},
\end{equation}
such as the field $E(t)=\expect{\wgmode{\text{out}}(t)}$, and
the output flux 
\begin{equation}\label{eq:power_output}
    P(t) = \expect{\wgmode{\text{out}}^\dagger(t) \wgmode{\text{out}}(t)}.
\end{equation} To monitor the coherence of the system, we also track the coherent and incoherent components of the output flux, $P(t) = P_{coh}(t) + P_{inc}(t)$, with $P_{coh}(t) = |E(t)|^2$. Finally, to study photon correlations and intensity fluctuations, we monitor the two-time second-order correlation $G^{(2)}(t_1,t_2)$ given by
\begin{equation}\label{eq:G2}
    G^{(2)}(t_1,t_2) = \expect{\wgmode{\text{out}}^\dagger(t_1) \wgmode{\text{out}}^\dagger(t_2)  \wgmode{\text{out}}(t_2)\wgmode{\text{out}}(t_1)},
\end{equation} and the normalized correlation
\begin{equation}\label{eq:g2}
    g^{(2)}(t_1,t_2) = \frac{G^{(2)}(t_1,t_2)}{P(t_1)P(t_2)}.
\end{equation} The two-time correlators can be computationally obtained via the quantum regression method \cite{gardiner2004quantum}. The application of the quantum regression method within the MF approach is detailed in Ref.~\cite {MFrobicheaux2021beyond}. In particular, to compute $g^{(2)}(0, t)$ for a system starting from the inverted state $\ket{\psi}_{N} = \allexcited$, we use the ratio of the power emitted from a lower Dicke state $\ket{\psi}_{N-1} \propto \sum_n \spindown{n} \ket{ee\dots e}$ to the power emitted from $\ket{\psi}_{N}$. This quantity measures the initial shot-to-shot fluctuation of the output photon rate \cite{bach2024emergence} starting from full inversion.


Because of the exponential computational complexity $(\mathcal{O}(4^N))$ of solving \Eq{\ref{eq:master}} and the lack of any simplifying symmetries in the chiral system \cite{MFchiral2023higherorder}, we use approximation techniques as described in the rest of this section. We first lay out the mean field or the cumulant expansion approach then turn to analytical methods based on a perturbative expansion.
\subsection{Mean Field Equations}\label{sec:mf_equation}

In this section, we present the mean field approach for the system in \Eq{\ref{eq:master}}. First, the Langevin-Heisenberg equation can be derived for a general atomic operator $\hat A$ resulting in \cite{TWA2024cascaded}

\begin{equation}\label{eq:langevin_general_op}
    \frac{\diff}{\diff t} \hat A =
    \sum_n  \left\{   [\spinup{n}, \hat A] \left(\frac{1}{2} \spindown{n} + \imu \inmodeatom{n} \right)  +  \left(\frac{1}{2}\spinup{n} - \imu (\inmodeatom{n})^\dagger \right)  [\hat A, \spindown{n}] \right\},
\end{equation} where $\inmodeatom{n} = \sqrt{\beta_n}\wgmode{n}$ is the input mode of the atom (assuming a vaccum state for the freespace field \cite{TWA2024cascaded}) and
\begin{equation}\label{eq:wgmode_at_n}
    \wgmode{n} = \alpha - \imu \sum_{m<n} \sqrt{\beta_m} \spindown{m},
\end{equation} where the sum is only over upstream atoms from the point of view of atom $n$, due to the unidirectionality of the waveguide. A similar equation for a bidirectional waveguide can be derived and is shown in App.~\ref{sec:bidirectional}.

The evolution of single atom operators ($\spindown{n}$ and $\exc{n}$) is obtained by substitution in \Eq{\ref{eq:langevin_general_op}} yielding
\begin{equation}\label{eq:langevin_sigma}
    \frac{\diff\spindown{n} }{\diff t} = -\frac{1}{2}\spindown{n} -\imu (1-2\exc{n}) \sqrt{\beta_n}\wgmode{n}, 
\end{equation} and 
\begin{equation}\label{eq:langevin_exc}
    \frac{\diff\exc{n} }{\diff t} = -\exc{n} -\imu \sqrt{\beta_n} \left(\wgmode{n} \spinup{n}- \wgmodedag{n}\spindown{n} \right), 
\end{equation}

These Heisenberg equations for operators are as hard to solve as the original master equation in Eq.~(\ref{eq:master}). However, we can turn them into complex-number equations by evaluating the expectation value with respect to the initial state $\rho(t=0)$ yielding, for example,
\begin{equation}\label{eq:langevin_sigma_expect}
    \frac{\diff\expect{\spindown{n}} }{\diff t} = -\frac{1}{2}\expect{\spindown{n}} -\imu \sqrt{\beta_n} (\expect{\wgmode{n}}-2\expect{\exc{n}\wgmode{n}}).
\end{equation} Note that the evolution of the expectation value of single atom operators requires the knowledge of the expectation value of two-atom operators. At this point, we can approximate the two-atom operators using a cumulant expansion \cite{CEkubo1962generalized, MFspinskramer2015generalized}
\begin{equation}\label{eq:cumulant2}
    \expect{\expect{\hat A_m \hat B_n}} \equiv \expect{\hat A_m \hat B_n} - \expect{\hat A_m } \expect{\hat B_n} \approx 0,
\end{equation} resulting in a closed ordinary differential equation (ODE) system for the single-atom operators. This method is the so-called mean-field or MF1 \cite{MFrobicheaux2021beyond}. MF1 ignores all correlations between the atoms and simply assumes that the system stays in a product state. Although this method is straightforward to use and is ubiquitously utilized in many-body physics, it is known to fail for the transient dynamics of a spin system near complete inversion \cite{PRXsuperradiant2024bursts,MFrobicheaux2021beyond,yelinPhysRevResearch.5.013091,MFDicke202509.1285}. This is because it fails to capture the buildup of correlations that are initially absent in an inverted system. For example, it predicts that an inverted system exponentially decays with no superradiant burst. Thus, in the remainder of the paper, we will not study this method. Rather, we will study a variant of MF1 from Refs.\cite{PRLLiedl2023collective,PRXsuperradiant2024bursts}.


In Refs.\cite{PRLLiedl2023collective,PRXsuperradiant2024bursts}, MF1 is known as the coherent state approximation (CSA), as it implies that a field mode $\wgmode{n}$ driving atom $n$ is a coherent field. In \cite{PRXsuperradiant2024bursts, ThesisLiedl2023collective}, the CSA was heuristically augmented by adding an incoherent part to the coherent mode $\wgmode{n} \rightarrow \wgmode{n\text{,coh}} + \wgmode{n\text{,inc}}$, yielding the mixed coherent state approximation (MCSA). MCSA can capture the buildup of correlations required for the superradiant burst and was successfully used to reproduce the first order correlator $P(t)$ in various experimental settings \cite{PRXsuperradiant2024bursts}. However, this method was not capable of obtaining the higher order correlator $g^{(2)}$. In this paper, for the sake of comparison with higher MF methods, we show some results of MCSA, as well as a heuristic extension of it to capture $g^{(2)}(0,t)$ starting from the inverted state.

The other alternative to MF1 (and MCSA) is to augment the ODE system with equations for two-atom operators, which will generally depend on 3-atom operators that can be approximated with another higher order cumulant expansion \cite{MFspinskramer2015generalized,MFrobicheaux2021beyond}. This is denoted as MF2 \cite{MFrobicheaux2021beyond}. This hierarchy of approximation can be extended if needed to capture higher order correlations. However, this comes at the expense of the size of the ODE system growing as $\sim (N^n)$, where $N$ is the number of atoms, and $n$ is the order of MF-$n$. In this paper, we investigate up to 3rd order mean-field (MF3) to capture relevant physics of interest including the emitted power and the second order coherence.

Similar to equations of single atom-operators Eq.~(\ref{eq:langevin_sigma}), the equations of motion for two or three atom operators required for higher order MF can also be derived. For example, the atom-atom correlator $\spinup{l}\spindown{k}$ evolves as, 
\begin{equation}\label{eq:langevin_sigmaplussigmaminus}
    \frac{\diff\spinup{l}\spindown{k} }{\diff t} = -\spinup{l}\spindown{k}  -\imu \sqrt{\beta_l} \wgmodedag{l}(2\exc{l} - 1)\spindown{k} + \imu \sqrt{\beta_k} (2\exc{k} - 1)\spinup{l}\wgmode{k} , 
\end{equation} and the rest of the MF2 equations is in App.~\ref{app:MF23} for clarity of the text.

The usage of the field operators $\wgmode{n}$ in the above equations has three purposes. First, it serves as a notational device to compactify the equations of motion. Otherwise, the equations would contain single sums over atoms. Second, it unravels the collective nature of the problem; the properties of a certain group of atoms depend only on the \emph{collective modes} of upstream atoms and not directly on the behavior of \emph{individual} atoms.
Third, it paves a route towards computational speedups. For a chiral system, such field operator at different atoms can be evaluated recursively as 
\begin{equation} \label{eq:wgmode_recursive}
    \wgmode{n} = \wgmode{n-1} - \imu \sqrt{\beta_{n-1}} \spindown{n-1},
\end{equation} and so can the field correlations. For example, in Eq. (\ref{eq:langevin_sigmaplussigmaminus}), the term $\spinup{l}\wgmode{k}$ is part of the MF2 equations and can be built up incrementally as
\begin{equation} \label{eq:wgmodecorr_recursive}
\spinup{l}\wgmode{k} = \spinup{l}\wgmode{k-1} - \imu \sqrt{\beta_{k-1}} \spinup{l}\spindown{k-1},
\end{equation} except when $l=k-1$, for which case $\spinup{l}\wgmode{k} = \spinup{l}\wgmode{k-1} - \imu \sqrt{\beta_{k-1}} \exc{k-1}$. By first computing the mode operator, $\expect{\wgmode{n}}$, and its correlation with the atomic operators iteratively, the derivatives in the equations of motion can be evaluated in $\mathcal{O}(N^n)$ steps instead of $\mathcal{O}(N^{n+1})$ for a generic extended ensemble \cite{MFspinskramer2015generalized} (for example in freespace coupled atoms). This allows us to simulate systems of large size $N \sim 10^4$ atoms using MF2 (see simulations in the companion paper \cite{tohfathermlimit}) or $N \sim 10^3$ atoms using MF3. Such system sizes are becoming accessible in recent experiments. Note that a similar speedup can be utilized even in the case of a bidirectional waveguide as illustrated in App.~\ref{sec:bidirectional}.
\subsection{Analytical $\beta$ expansion} \label{sec:analytical_methods}
In this section, we outline a method to extract analytical solutions in special cases. It utilizes the small coupling $\beta$ (as in the experiment) to perform a Taylor expansion of the desired observables. The advantage of this analytical method over other methods is that it can be extrapolated to large $N$ in a relatively easy way. This allows us to explore the large $N$ behavior as well as benchmark the MF method at such high $N$. The cases considered do not include an input field, i.e., $\alpha(t) = 0$. To allow for scaling, the atoms also start in a symmetric state and we assume a homogeneous coupling, $\beta$, for all atoms. 

Here, we consider starting in the excited state, $\ket{ee\dots e}$, or a lower Dicke state, $\ket{\psi}_{N-1} \propto \sum_n \spindown{n} \ket{ee\dots e}$, and we track either the radiated power, $P(t)$, or the second order correlator, $G^{(2)}(t,t)$. This information is sufficient to compute $g^{(2)}(t,t)$ (as done using the TWA method in Ref.~\cite{TWA2024cascaded}) or $g^{(2)}(0,t)$ as measured by the experiment in Ref.~\cite{bach2024emergence} starting from the excited state, $\allexcited$. The details of this method are in App.~\ref{sec:analytical_methods_details}. The end goal is to write the aforementioned observables as a series expansion in powers of $\beta$. For example, if the system is initialized in the excited state, $\ket{ee\dots e}$, the radiated power, $P$, in \Eq{\ref{eq:power_output}} is

\begin{align}\label{eq:analytical_power}
    P(t, N, \beta) &= \sum_{k=1} \beta^k f_k(N,t),
\end{align} where
\begin{align}
    f_1(N, t) &= N e^{-t},
\end{align}
\begin{align}
    f_2(N, t) &= -N (N - 1) \left[ 2 e^{-2t} + (-2 + t)e^{-t} \right],
\end{align}
\begin{align}
    f_3(N, t) &= \frac{1}{2} N (N - 1)^2 (t - 2)^2 e^{-t} \nonumber \\ &+ 2N (N - 1)(N - 2) t e^{-2t} \nonumber \\ &- 2N (N - 1)(2N - 3)e^{-2t} \nonumber \\ &+ 2N (N - 1)(N - 2)e^{-3t}.
\end{align} The successive functions become increasingly intricate. Here, we only give the first few coefficients. Because of computational difficulties outlined in App.~\ref{sec:analytical_methods_details}, we usually stop the expansion at $k_{max}$ around 7 or 8. For the sake of illustrating convergence with the truncation order $k_{max}$, we will use the highest two available orders in the results section Sec.~\ref{sec:results}.

The first order contribution $f_1(N, t)$ is a non-cooperative contribution to the power radiated into the waveguide and is the only contribution predicted by MF1. Generally, the $\nth{k}$ order contribution comes with a polynomial in $N$ of degree $k$. This means for finite $N$, the dynamics depends on both $N$ and $\beta$ individually but in the large $N$ limit, the dynamics is a function of only $N\beta$, which serves as the new expansion parameter. Also, $f_k$ has vanishing low-order initial time derivatives, i.e. $\partial^m_t f_k(N, t)|_{t=0} = 0$ for $m < k - 1$, meaning the higher orders contribute only at larger $t$. Additionally, we note the multi-timescale nature of the solution represented by exponentials with multiple decay rates as well as polynomials in $t$ with increasing degrees. These multiple timescales come from the quantum state successively traversing bigger portions of the Hilbert space. They are also reminiscent of the intricate dynamics of a chiral system without simplifying symmetries.

The analytic $\beta$ expansions such as in \Eq{\ref{eq:analytical_power}} have intuitive convergence characteristics. First, they converge faster for early time than for late time, showing that more terms are needed as time progresses and several time-scales start to compete. Second, they converge faster for small $N \beta$ values than for large values. This is natural as a larger value of $\beta$ means a stronger perturbation from the trivially exact first-order contribution in the $\beta \rightarrow 0$ limit. Additionally, larger $N$ means more complexity coming from the propagation of the dynamics through more atoms. Finally, for a fixed $N \beta$, the series expansion converges faster for bigger $N$. In fact, the dynamics of some observables exhibits a simple closed form in the limit $N \rightarrow \infty$, and MF2 becomes exact, agreeing with the analytical solution. This $N \rightarrow \infty$ limit is explored in the companion paper \cite{tohfathermlimit}. Similar analysis can be done for $G^{(2)}(t,t)$ starting from full inversion as well as for $P(t)$ starting from the $\ket{\psi}_{N-1}$.

\section{Results} \label{sec:results}
In this section, we present results regarding several quantities pertaining to photon statistics in the chiral waveguide system. First, we benchmark the MF approach presented in Sec.~\ref{sec:mf_equation} for small $N$ by comparing to exact calculations and discuss the behavior of MF2 and MF3. Second, we benchmark the MF approach for relatively large $N$ by comparing to analytical results presented in Sec.~\ref{sec:analytical_methods}. Third, we discuss the simulation of the experimental results presented in Fig.~\ref{fig:intro_figure} and show large sensitivity to the excitation pulse area. Finally, we establish a limitation of the low-order MF approach regarding the emergence of second order coherence starting from idealized inversion.

For the experimental simulations, we work in the SI units. We use the D2 transition of nanofiber-coupled cold cesium atoms. The excited state lifetime ($1/\Gamma_0$) is 30.5~ns. The transition frequency is $ \omega_0/(2\pi) = 3.52 \times 10^{14}$ ~Hz, with a wavelength of $852$~nm. For the simulations that compares the different approximation methods, we work with the dimensionless quantities, where time is measured in units of $1/\Gamma_0$, and the power is measured in units of $\Gamma_0\hbar\omega_0$.

\subsection{Benchmarking the MF approach at small $N$}
\label{sec:benchmarking_small_n}

To establish the validity of the MF methods for a chiral waveguide, we benchmark the MF approach in various settings. The first is in the small $N$ regime which can be simulated exactly using the DM or MC trajectories. Motivated by the experiment in Ref.~\cite{bach2024emergence}, in most of our tests, we fix the optical depth $\propto \beta N \sim 10$. We varied the optical depth in different tests (not shown here), and as expected, it was found that the accuracy of the various approximations decreases with increasing optical depth.





\begin{figure}[htbp]
    \centering
    \includegraphics[width=0.45\textwidth]{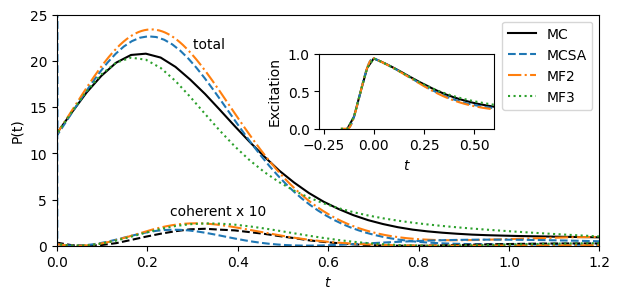}
    \caption{\justifying Comparison of the radiation and excitation dynamics for the MC, MCSA, MF2, and MF3 approaches with $N = 18$ atoms and $\beta = 0.5$ starting from the ground state. The total power emitted into the waveguide, $P(t)$, after the pulse and the small coherent component, $P_{coh}(t)$, are shown. $P_{coh}(t)$ is multiplied by $10$ for clarity. The inset shows the excitation during and after the pulse.}
    \label{fig:mf_comparison_N18}
\end{figure}

In Fig.~\ref{fig:mf_comparison_N18}, we set $N = 18,\, \beta = 0.5$ starting from the ground state $\ket{gg \dots g}$, and we send a square pulse of duration $4$ ns (from $t= -4$ ns to $t = 0$ ns) in the waveguide. This is a short pulse motivated by the experiment and has pulse area, $A = 1.13 \pi$, which maximally inverts the system (see Sec.~\ref{sec:simulation_experiment} and Ref.~\cite{PRXsuperradiant2024bursts} for the interpretation of the quantity $A$). After the pulse ($t \geq 0$), we record the subsequent power emitted into the waveguide. We find the MF2 and MF3 approximations giving qualitatively correct behavior. We also compare to the MCSA from \cite{PRXsuperradiant2024bursts}, which we find to agree closely with MF2 throughout most of the tests.

In Fig.~\ref{fig:mf_comparison_N18}, all methods quantitatively agree as the correlations buildup. As the maximum of the correlation and superradiance is approached, the different methods start to differ but still qualitatively agree, with MF3 generally being the closest to the exact calculation based on MC trajectories.
While the MF2 and MCSA overshoot the peak power, MF3 tends to underestimate the peak power. This is a general pattern we observe in various settings and is in accord with recent investigations \cite{MFDicke202509.1285}. In Ref.~\cite{MFDicke202509.1285}, the even orders of MF tend to overshoot the peak power, while the odd orders of MF undershoot. Thus, a good estimate of the actual power might come from an average of both odd and even orders.

Note that because the system is maximally inverted, most of the power comes out as an incoherent power with a small fraction emitted coherently. In the inset, we show the time dependent average atom excitation which agrees for all models during the short pulse and reaches $\approx 94 \%$ at the end of the pulse. The excitation then decays superradiantly. During the superradiant burst, the excitation is smaller for a method that has a larger peak power. Note that the radiation is partially synced at the end of the pulse because the cooperative enhancement $P(0) - \beta N \times \text{excitation}(0) = 30 \% \times P(0)$ is an appreciable fraction of the initially emitted power. This comes about from the method of excitation and the finite duration of the pulse. As the pulse is inverting the system over a finite time extent, the atoms start locking their radiation, such that at the end of the pulse, they are partially phase-locked.

\subsection{Benchmarking the MF approach at large $N$}
\label{sec:benchmarking_big_n}

\begin{figure}[htbp]
    \centering
    \includegraphics[width=0.45\textwidth]{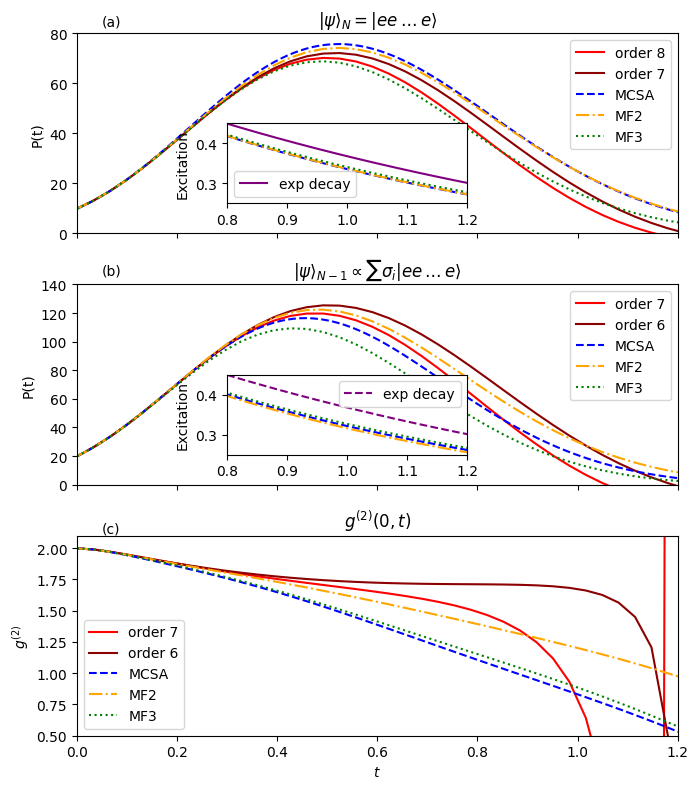}
    \caption
    {\justifying Comparison of the emission power into the waveguide, $P(t)$, and the second-order correlator, $g^{(2)}(0,t)$, for the analytical solutions truncated at $k_{max} = 6,7,8$, MCSA, MF2, and MF3 approaches with $N = 900$ atoms and $\beta = 0.011$. (a) $P(t)$ starting from the fully inverted state. (b) $P(t)$ starting from a lower Dicke state $\ket{\psi}_{N-1} \propto \sum_n \spindown{n} \ket{ee\dots e}$. (c) the ratio of the power from (b) to the power from (a) gives $g^{(2)}(0,t)$. The insets show the excitation fraction decaying with time.
    }
    \label{fig:mf_comparison_N900}
\end{figure}

In this section, we benchmark the MF and MCSA approaches against the analytical results obtained in Sec.~\ref{sec:analytical_methods} for large $N = 900$ and $\beta = 0.011$ starting from the fully inverted or a lower Dicke state,  $\ket{\psi}_{N-1} \propto \sum_n \spindown{n} \ket{ee\dots e}$, (required for $g^{(2)}(0,t)$). These parameters are motivated by the experiments in \cite{PRXsuperradiant2024bursts,bach2024emergence}. The results are shown in Fig.~\ref{fig:mf_comparison_N900}.

In the top panel, starting from the fully inverted state, the pattern of over- and underestimation is the same in Fig.~\ref{fig:mf_comparison_N18} with the MC calculation replaced by two consecutive orders of the analytical solution. In the middle panel, the MCSA undershoots like MF3. For the analytical solution, we note that the two consecutive orders only converge until about the peak power. After the peak power, they start to deviate giving a non-physical negative power at late times. This non-physical behavior does not happen with other methods. In the bottom panel, the ratio of the power from a lower Dicke state to the power from the top Dicke state gives $g^{(2)}(0,t)$. There, MF3 and MCSA behave similarly. The analytical solution only converges until about the peak power, and before this point it agrees with MF2 more than with other methods.

For the MCSA calculation, we follow the steps in Ref.~\cite{ThesisLiedl2023collective}. The technique thereof was established for the initial state $\allexcited$. Here, we extend this method to simulate the initial state $\ket{\psi}_{N-1} \propto \sum_n \spindown{n} \ket{ee\dots e}$. For the large atom number $N = 900$ considered, it can be shown that the collective state $\ket{\psi}_{N-1}$ is close to an ensemble of product states with the individual atom state $\sqrt{1 - 1/N}\ket{e} + e^{\imu \phi} \sqrt{1/N} \ket{g}$ after averaging (or having a convex combination) over the phase $e^{\imu \phi}$. This is an application of the quantum de Finetti theorem for symmetric Dicke states in the limit $N \rightarrow \infty$ \cite{deFenittikoenig2009most}. The pure state, $\ket{\psi}_{N-1}$, and this proposed ensemble have the same single atom expectation values ($\expect{\spindown{n}}$ and $\expect{\exc{n}}$), while they differ in the two atom expectation values ($\expect{\spinup{l}\spindown{k}}$ and $\expect{\exc{l}\exc{k}}$) by a factor that scales as $\sim \mathcal{O}(\frac{1}{N^2})$, justifying the approximation for large $N$. For the MCSA simulation in the middle panel of Fig.~\ref{fig:mf_comparison_N900}, we used this approximate ensemble and averaged over many values of $\phi$ until reaching convergence. 

\subsection{Simulation of the experiment} \label{sec:simulation_experiment}
In this section, we discuss details of the simulation that was used to reproduce the experimental measurements of Ref.~\cite{bach2024emergence} in Fig.~\ref{fig:intro_figure}. We also discuss the sensitivity to the inversion pulse area. In this simulation, there are $N = 900$ atoms which start in the all ground state $\ket{gg\dots g}$. Due to thermal fluctuations that result in the movement of atoms radially from the nanofiber, the atoms have fluctuating coupling strength, $\beta \sim p(\beta)$. The distribution $p(\beta)$ is a Gaussian of mean $0.0108$ and standard deviation $0.0051$ truncated so that $0 \leq \beta \leq 1$. This results in $\expect{\beta} = 0.01103$ and $\expect{\sqrt{\beta}} = 0.1019$, and the expected optical depth $ \propto \expect{N \beta} \approx 10$.

A pulse of duration $T_p = 4$ ns $\ll 1/\Gamma_0$ and a certain power, $P_p$, is applied to the waveguide from $t = -4$ ns to $t = 0$ ns. This results in a mean pulse area seen by the first cesium atom as \cite{PRXsuperradiant2024bursts}
\begin{equation}
    \expect{A} \equiv \sqrt{\frac{4\Gamma_0 P_p}{\hbar\omega_0}} T_p \expect{\sqrt{\beta}}.
\end{equation}
As shown in Ref.~\cite{bach2024emergence}, the mean pulse area, $\expect{A}$, required for maximal inversion is at $\expect{A} = 1.07 \pi$. The difference from $\pi$ comes about from three competing factors. First, the finite time of the pulse tends to shift the optimal $\expect{A}$ to larger values compared to an infinitely short pulse. Second, the fluctuation of the coupling, $\beta$, tends to shift $\expect{A}$ to smaller values compared to an idealized homogeneous coupling. Third, the absorption of the pulse along the array depletes its power, and there is a subsequent need to compensate by a higher pulse power (area) \cite{PRXsuperradiant2024bursts,bach2024emergence}. This can be theoretically shown by simulations as was done by the TWA method in Ref.~\cite{bach2024emergence} and the MF1 method in Refs.~\cite{PRXsuperradiant2024bursts,PRLLiedl2023collective} and verified in this work by the MF2 approach.

We model the short pulse as measured by the experiment. The pulse has rising and falling edges that have a characteristic time scale of $\sim 0.5$ ns. To characterize the photon statistics from the superradiantly decaying ensemble, we make sure that the excitation pulse has become negligible and start recording at $t \approx 1$ ns for the observables in Fig.~\ref{fig:intro_figure}.


While the pulse area required for maximal inversion is $\expect{A} = 1.07 \pi$, the data in Fig.~\ref{fig:intro_figure} was collected at $\expect{A} = 1.1 \pi$. The average excitation (see inset) present at the switch off of the excitation pulse ($t = 1$ ns) is $80 \%$ and the coherent power represents about $ 50\%$ of the total radiated power. This is a signature that the system is partially coherent with the incident laser field after the pulse has ended \cite{PRLLiedl2023collective}. This coherence is only lost at maximal inversion at $A = 1.07 \pi$ where the coherent radiation represents only a tiny fraction of the total initial power.

\begin{figure}[h]
    \centering
    \includegraphics[width=0.4\textwidth]{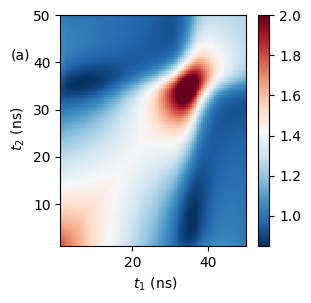} 
    \includegraphics[width=0.4\textwidth]{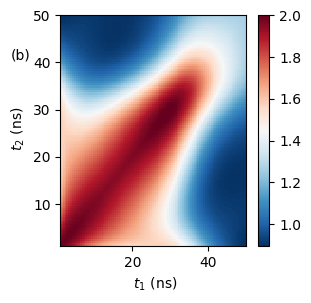} 
    \caption{\justifying Two-time second order correlation $g^{(2)}(t_1,t_2)$ using the MF2 method. (a) near maximal inversion using a pulse area $A = 1.1 \pi$. (b) at maximal inversion using a pulse area $A = 1.07 \pi$. }
    \label{fig:g2_correlation}
\end{figure}

For the experimentally available data at $\expect{A} = 1.1 \pi$, we verify the agreement of MF2 for the two-time correlator $g^{(2)}(t_1,t_2)$ at $(t_1,t_2) = (4.5\, \text{ns},t)$, which is shown in Fig.~\ref{fig:intro_figure}. In addition, we verify the agreement for the same time $(t_1,t_2) = (t,t)$. We provide the full two-time second order correlation $g^{(2)}(t_1,t_2)$ in Fig.~\ref{fig:g2_correlation} as predicted by MF2. We note the initial second order coherence $g^{(2)}(1 \text{ns},1 \text{ns})$ is $1.8$ for $A = 1.1 \pi$, which is deviant from that of a maximally inverted state at an inversion pulse area $A = 1.07 \pi$, where $g^{(2)}(1,\text{ns},1\, \text{ns}) \approx 2$ (see Fig.~\ref{fig:g2_correlation}(b)). 

For maximal inversion, we do not have experimentally available data. In this case, MF2 predicts that the same time coherence $g^{(2)}(t,t)$ is $\approx 2$ for $ t < 30$ ~ns. This shows a \emph{lack} of build-up of second order coherence at \emph{maximal} inversion, which is expected if the system starts from \emph{perfect} inversion. Starting from the perfectly inverted state, $\allexcited$, $g^{(2)}(t,t)$ is expected to drop from $2$ (an independently radiating ensemble) to $\sim 1$ (a coherently radiating ensemble) when the atoms have homogeneous coupling, $\beta = 0.01$, according to the TWA method in Ref.~\cite{TWA2024cascaded}.

This apparent discrepancy raises a question about the reliability of the MF2 calculation at maximal inversion especially for the same-time correlator $g^{(2)}(t,t)$. In Sec.~\ref{sec:coherence_full_inverted}, we show that MF2 fails to predict $g^{(2)}(t,t)$ starting from \emph{perfect} inversion, and illustrate the need for MF4. However, at \emph{maximal} inversion  using the laser pulse, it is not clear to us whether the prediction in Fig.~\ref{fig:g2_correlation}(b) is accurate or not.



\subsection{Second order coherence at full inversion} \label{sec:coherence_full_inverted}

In this section, we describe a limitation of low order mean field approximations. While capable of capturing the second order coherence away from maximal inversion, MF2 and MF3 are unable to capture the buildup of second order coherence when starting from the fully inverted state $\allexcited$, which is predicted by the TWA method \cite{TWA2024cascaded} as well as our analytical results. We show that a higher order mean field (MF4 or higher) is required to capture the same-time correlator $g^{(2)}(t,t)$ for perfect inversion. While we have not implemented the MF4 calculation for the chiral system under investigation due to the relative complexity of the equations of motion and the expensive compute time ($\mathcal{O}(N^4)$), we compare the MF2 and MF3 to the analytically available solution for homogeneous $\beta$. In App.~\ref{sec:continuum_approximation_sym}, we show that a MF4 or higher method can reliably predict $g^{(2)}(t,t)$ for a \emph{permutationally symmetric} system, which has a much simpler ODE system.

Our analytical solution to $g^{(2)}(t,t)$ in \Eq{\ref{eq:g2}} comes from approximating both its numerator, $G^{(2)}(t,t)$, in \Eq{\ref{eq:G2}} and denominator. In the following, we use the highest available order for the denominator, while varying the truncation order of the numerator to check convergence. We use moderate parameters: atom number $N = 150$ or $N = 300$ and coupling $\betaUni = 0.01$, which have reasonable convergence for $t < 1.5$. Another reason for this choice is that we can compare with the TWA approximation in Ref.~\cite{TWA2024cascaded}.

\begin{figure}[htbp]
    \centering
    \includegraphics[width=0.45\textwidth]{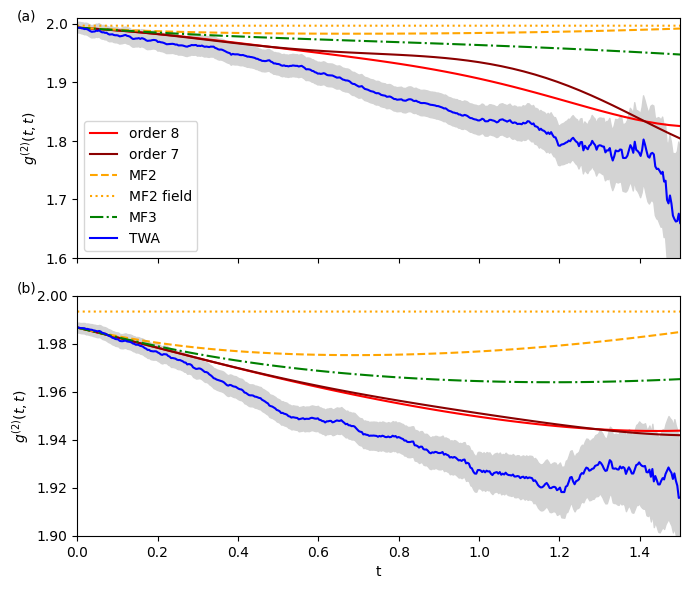}
    \caption{\justifying Comparison of the dynamics of $g^{(2)}(t,t)$ starting from the totally inverted state via MF, analytical approximations of truncation orders $7$ and $8$, and the TWA method from Refs.~\cite{TWA2024cascaded,bach2024emergence}. The cumulants in MF2 were done both at the `standard' atomic level (denoted as `MF2'), and at a `field' level (denoted as `MF2 field'). The gray shaded area represents the statistical uncertainty of the TWA data. Parameters are $\betaUni = 0.01$ and $N=300$ (a) or $N=150$ (b)}
    \label{fig:mf_comparison_g2tt}
\end{figure}


In Fig.~\ref{fig:mf_comparison_g2tt}(a), the analytical solution of order $8$ shows a drop from $ \approx 2$ to $\approx 1.91$ at $t \approx 1$. We believe this order of approximation to be accurate until $t \approx 1$. We note that the TWA tends to overestimate the drop in $g^{(2)}(t,t)$ as it falls from $ \approx 2$ to $\approx 1.83 \pm 0.025$ at $t \approx 1$ (in units of the lifetime $1/\Gamma_0$). To confirm this pattern, we show a case with a smaller optical depth, where $N = 150$ and $\beta = 0.01$, in Fig.~\ref{fig:mf_comparison_g2tt}(b). There, the analytical solutions of order $7$ and $8$ are convergent in the time window in Fig.~\ref{fig:mf_comparison_g2tt} and show a drop from $ \approx 1.985$ at $t = 0$ to $ \approx 1.945$ at $ t = 1.2$. On the other hand, the TWA predicts a drop to $\approx 1.92 \pm 0.01$ at $t = 1.2$.

In Fig.~\ref{fig:mf_comparison_g2tt}, the MF2 and MF3 methods struggle to show such drop of $g^{(2)}(t,t)$ compared to the analytical solution and the TWA. An intuitive reason for this failure can be attributed to the cumulant expansion used in MF2 and MF3 starting from full inversion as follows.
 
The output fields in the second order correlator $G^{(2)}(t,t) = \expect{\wgmode{\text{out}}^\dagger(t) \wgmode{\text{out}}^\dagger(t)  \wgmode{\text{out}}(t)\wgmode{\text{out}}(t)}$ are composed of individual atomic operators as in \Eq{\ref{eq:output}}. The standard treatment for the cumulant expansion is done at the atomic level by first substituting \Eq{\ref{eq:output}} and then doing the cumulant expansion (see Ref.~\cite{MFrobicheaux2021beyond}). In the large $N$ cases, however, the cumulant expansion can be done at a macroscopic or a field level without much loss in accuracy \cite{MFDicke202509.1285}. In the transient dynamics starting from full inversion, only correlators with same number of raising and lowering field operators are non-zero. For example, $\expect{\wgmode{\text{out}}(t)} = \expect{\wgmode{\text{out}}^\dagger(t) \wgmode{\text{out}}^\dagger(t)  \wgmode{\text{out}}(t)} = 0$. A cumulant expansion at a MF2 or MF3 then reveals that  
\begin{align}\label{eq:G2tt_cumulant}
    G^{(2)}(t,t) &= 2 \expect{\wgmode{\text{out}}^\dagger(t)  \wgmode{\text{out}}(t)}^2 \nonumber \\
    &+ \expect{\wgmode{\text{out}}^\dagger(t)  \wgmode{\text{out}}^\dagger(t)}\expect{\wgmode{\text{out}}(t)  \wgmode{\text{out}}(t)}  \nonumber \\
    &- 2 \expect{\wgmode{\text{out}}^\dagger(t)}^2 \expect{\wgmode{\text{out}}(t)}^2 = 2 (P(t))^2,       
\end{align} resulting in a trivial $g^{(2)}(t,t) = 2$. 


To validate this reasoning, we used a MF2 method based on an expansion at a `field' level resulting in $g^{(2)}(t,t) \approx 2$, which is shown in Fig.~\ref{fig:mf_comparison_g2tt}. Even at the standard microscopic expansion, MF2 and MF3 predict that $g^{(2)}(t,t)$ stay close to $2$, deviating appreciably from the analytical prediction. The failure of the cumulant expansion at both MF2 and MF3 reveals that at least 4-body correlations are necessary for the proper behavior of a $g^{(2)}(t,t)$. In general, this analysis reveals that a higher order MF is required for higher order correlators, especially when coherent dynamics is absent ($\expect{\spindown{i}(t)} = \expect{\spindown{i}(t)\spindown{j}(t)} = 0$ as in starting from perfect inversion). We show in App.~\ref{sec:continuum_approximation_sym} that a MF4 approximation (which does not use a cumulant expansion for 4-body correlators) is in good agreement with the exact result based on a density matrix calculation for a permutationally symmetric system with parameters $\beta = 0.01$ and $N = 300$.


Despite the failure of MF2 and MF3 to reliably predict $g^{(2)}(t,t)$ at full inversion, MF2 is able to predict all experimentally available data in Ref.~\cite{bach2024emergence}, which is close to, but not at, maximal inversion. In the case of maximal inversion, the atoms in a chiral system have appreciable dipole moments, $\expect{\spindown{i}}$ as shown in Ref.~\cite{PRXsuperradiant2024bursts}. Such dipole moments may seed coherent dynamics within the array, resulting in a non-vanishing output field, $E(t) = \expect{\wgmode{\text{out}}(t)}$. The presence of a finite coherent field renders the cumulant expansion at the MF2 (as in \Eq{\ref{eq:G2tt_cumulant}}) or MF3 level more involved than in the case of perfect inversion, complicating analytical reasoning about $G^{(2)}(t,t)$. Consequently, under experimentally relevant conditions of maximal inversion achieved via an optimal waveguide pulse, it remains unclear whether the MF2 prediction shown in Fig.~\ref{fig:g2_correlation}(b) is quantitatively reliable.



\section{Conclusion}
\label{sec:conclusion}
In this work, we presented two approaches to study photon statistics in WQED focusing on a chirally coupled array of atoms which recent experiments have demonstrated. Using a higher order mean field method allowed us to reproduce recent experimental results in Ref.~\cite{bach2024emergence}. Although a higher order mean field approach is generally expensive to compute, we presented a method to speed up the computations by a factor of $N$, allowing us to simulate up to tens of thousands of atoms using MF2 (as done in the companion paper \cite{tohfathermlimit}).

The second approach relies on systematically solving the dynamics starting from symmetric states and extrapolating to a general $N$. This allowed us to access converging results for moderate atom numbers ($\sim 1000$) with a moderate coupling constant. In addition, it exposed the complicated multi-time scale nature of a symmetry-lacking system. Finally, we demonstrated a possible shortcoming of MF2 and MF3 regarding the emergence of second order coherence in an \emph{ideally inverted} array. There, a MF4 or higher method is required for a proper buildup of second order coherence.


Various questions for future investigation emerge from our the results above. First, as we alluded to in the main text, it is not clear if a MF2 method is sufficient for photon-photon correlations in \emph{maximally inverted} array with an optimal short pulse as is done in the experiment. It is of interest to explore this question both theoretically and experimentally by using a pulse area aiming for \emph{maximal inversion}. Second, our analytical methods for finite $N$ were obstructed by a computationally expensive symbolic matrix power. We usually stopped at a matrix power (corresponding to a highest order in a perturbative expansion) of $\sim 10$. It is of interest to explore ways that could push this computational bottleneck to higher orders. This would allow for converging results for a wider range of experimentally relevant parameters.

Data plotted in the figures is available at \cite{data}.

\begin{acknowledgments}
    We thank Constanze Bach for providing data and details of the experiment in \cite{bach2024emergence}. 
    We thank Philipp Schneeweiß and Christian Liedl for useful discussions regarding the experimental setup in Ref.~\cite{PRXsuperradiant2024bursts}. We thank Saptarshi Saha for discussions of the MF approach. We thank Wenqi Tong and AbdAlghaffar Amer for useful discussions and suggestions.

    This work was supported by the National Science Foundation under Award No. 2410890-PHY. This research was supported in part through computational resources provided by Information Technology at Purdue University, West Lafayette, Indiana.
\end{acknowledgments}


\appendix
\section{Bidirectional WQED}\label{sec:bidirectional}
In this section, we show that the Heisenberg Eq. (\ref{eq:langevin_general_op}) can be extended to a bidirectional waveguide. If the $\nth{i}$ atom emits with rates $\betaf_i$ and $\betab_i$ in the forward and backward directions respectively, one can define the single atom constants $r_i = \sqrt{\betaf_i} e^{-\imu k_{1D} z_i}$ and $l_i = \sqrt{\betab_i} e^{+\imu k_{1D} z_i}$, where $k_{1D}$ is the wavenumber of the transition. Similar to the right propagating mode $\wgmode{i}$, one can also define a similar left propagating mode, $\wgbmode{j}$
\begin{equation}
    \wgmode{j} = \alpha_f - \imu \sum_{i<j} r_i \spindown{i},
\end{equation}
\begin{equation}
    \wgbmode{j} = \alpha_b - \imu \sum_{i>j} l_i \spindown{i},
\end{equation} where $\alpha_f$ and $\alpha_b$ are the forward and backward input fields.

The master equation in this case can be written compactly as

    \begin{align} \label{eq:master_bidirectional}
        \ddt \density &= -\imu \sum_n \{ r_n^* [\spinup{n}, \wgmode{n} \density] + r_n [\spindown{n}, \density \wgmodedag{n}] + l_n^* [\spinup{n}, \wgbmode{n} \density] \\ &+ l_n [\spindown{n}, \density \wgbmodedag{n}]\} + \mathcal{L}_{ng}[\density],   
    \end{align}
 where $\mathcal{L}_{ng}[\density] = \sum_n  \spindown{n} \density \spinup{n} - \frac{1}{2} \left\{ \spinup{n}\spindown{n}, \density \right\}$. From this equation, one can obtain the generalization of the Heisenberg Eq. (\ref{eq:langevin_general_op})

\begin{align} \label{eq:langevin_bidirectional_general_op}
    \frac{\diff}{\diff t} \hat A &=
    \sum_n \{[\spinup{n}, \hat A] \left(\frac{1}{2} \spindown{n} + \imu r_n^* \wgmode{n}+ \imu l_n^* \wgbmode{n} \right)   \nonumber \\  
    & \quad +  \left(\frac{1}{2}\spinup{n} - \imu r_n \wgmodedag{n} - \imu l_n \wgbmodedag{n} \right)  [\hat A, \spindown{n}]\}.        
\end{align}
This way, in order to speed up computations, one has to keep track of both the forward and backward modes and their correlations with the atomic operators and compute them recursively before evaluating the MF derivatives.

\section{MF2 Equations}\label{app:MF23}
In this section, we present the MF2 equations for the chiral (unidirectional) system in Eqs.~(\ref{eq:cascaded_superoperators}). The first step is to derive the Heisenberg equation for two-atom operators. 

\begin{equation}\label{eq:langevin_sigmaplussigmaminus_app}
    \frac{\diff\spinup{l}\spindown{k} }{\diff t} = -\spinup{l}\spindown{k}  -\imu \sqrt{\beta_l} \wgmodedag{l}(2\exc{l} - 1)\spindown{k} + \imu \sqrt{\beta_k} (2\exc{k} - 1)\spinup{l}\wgmode{k}, 
\end{equation}
\begin{equation}\label{eq:langevin_sigmaminussigmaminus}
    \frac{\diff\spindown{l}\spindown{k} }{\diff t} = -\spindown{l}\spindown{k}  + \imu \sqrt{\beta_l} (2\exc{l} - 1)\spindown{k}\wgmode{l} + \imu \sqrt{\beta_k} (2\exc{k} - 1)\spindown{l}\wgmode{k}, 
\end{equation}
\begin{align}\label{eq:langevin_excsigmaminus}
    \frac{\diff\exc{l}\spindown{k} }{\diff t} &= -\frac{3}{2}\exc{l}\spindown{k}  + \imu \sqrt{\beta_l} \wgmodedag{l}\spindown{l}\spindown{k} \nonumber \\ 
    &- \imu \sqrt{\beta_l} \spinup{l} \spindown{k}\wgmode{l}  + \imu \sqrt{\beta_k} (2\exc{k} - 1)\exc{l}\wgmode{k}, 
\end{align}
    \begin{align}\label{eq:langevin_excexc}
        \frac{\diff\exc{l}\exc{k} }{\diff t} &= -2\exc{l}\exc{k} + \imu \sqrt{\beta_l} \wgmodedag{l}\spindown{l}\exc{k} - \imu \sqrt{\beta_l} \spinup{l} \exc{k}\wgmode{l}  \nonumber \\&+ \imu \sqrt{\beta_k} \wgmodedag{k}\spindown{k}\exc{l} - \imu \sqrt{\beta_k} \spinup{k} \exc{l}\wgmode{k}.       
    \end{align}
Note that in the above equations, $\wgmodedag{m}$ always appears as the far left operator in products of operators, while $\wgmode{n}$ always appears to the far right. This is a rule of thumb to keep track of the order of the operators. The other operators have different indices and thus always commute. The mean field equation are then obtained by taking the expectation value and using the cumulant expansion to turn a 3-body operator into two and single body operators \cite{MFrobicheaux2021beyond}. As discussed in Sec.~\ref{sec:mf_equation}, in order to speed up the mean field computation by a factor of $N$, one has to keep track of the field $\expect{\wgmodedag{m}}$, as well as the field-atom correlators $\expect{\spinup{l}\wgmode{k}}$, $\expect{\spindown{l}\wgmode{k}}$, and $\expect{\exc{l}\wgmode{k}}$, and build them recursively before evaluating the MF derivatives. The MF3 equations are similarly derived but are more complicated and are omitted for brevity.

\section{Details of the analytical $\beta$ expansion} \label{sec:analytical_methods_details}

In this section, we describe the details of the analytical $\beta$ expansion in Sec.~\ref{sec:analytical_methods}. The method starts with writing the equations of motion in the Fock-Liouville space for a small $N \lessapprox 10 $ atoms. This is an exact remapping of the master equation and does not include any approximation. In this picture, the density matrix, $\density$, is reshaped into a vector, $\vec{\rho}$, and the Lindblad superoperator can be mapped into a matrix, $M$, which is called Liouvillian matrix. As a result, the master equation becomes a matrix-vector product form:


\begin{equation}\label{eq:linear_system}
    \frac{\diff \vec{\rho} }{\diff t} = M \vec{\rho} = (- M_0 - \beta M_1) \vec{\rho},
\end{equation} where $M$ is decomposed into a diagonal matrix, $M_0$, that dictates that all states are exponentially decaying and comes from the individual decay of the atoms in \Eq{\ref{eq:collectivedecay}}. $M_1$ couples different states and comes with strength $\beta$. Due to the special nature of the chiral system noted in $\cite{MFchiral2023higherorder}$, the matrix $M_1$ has some special properties. It could be visualized as an adjacency matrix of a weighted directed graph whose nodes are the states of the Fock-Liouville space and whose directed edges are weighted by the matrix elements of $M_1$. The resulting graph is \emph{acyclic}, as the nodes do not have reciprocal dependencies that form loops. This property does not hold for a general bidirectional system (except at the permutationally symmetric mirror or Bragg configuration \cite{anaPhysRevLett.131.033605}, where the Dicke states form a downward cascade in adjacent ladders, see \cite{permsymshammah2018open}). In principle, the acyclic property allows for an iterative elementary solution (with no need to solve an eigenvalue problem that comes from cycle dependencies) as was done in the single-excitation subspace \cite{oscillationPhysRevLett.128.203601}.



A formal analytical solution to Eq. (\ref{eq:linear_system}) can be obtained after a Laplace transform $\vec{\rho}(t) \rightarrow \vec{\Rho}(s)$

\begin{equation}\label{eq:linear_system_laplace}
    s \vec{\Rho}(s) - \vec{\rho}(0) = M \vec{\Rho}(s),
\end{equation}
which can be formally solved and expanded

\begin{align}\label{eq:linear_system_soln}
    \vec{\Rho}(s) &= (s - M)^{-1} \vec{\rho}(0) = (sI + M_0 + \beta M_1)^{-1} \vec{\rho}(0) \nonumber \\
    &= \sum_k \beta^k (A^{-1}M_1)^k A^{-1} \vec{\rho}(0),
\end{align} where $A = sI + M_0$ is a diagonal matrix, and the last sum is known as the Neumann series \cite{neumanwu2013approximate}. There are two reasons why we use the Neumann expansion. First, it reveals the collective interaction as successive powers of the interaction strength $\beta$. Second, it is computationally more feasible to find a truncated approximate of the inverse $(s - M)^{-1}$ than finding the full inverse. In principle, we can obtain the full solution because the sum truncates due to the acyclic property. We found for small $N$ cases that the sum truncates at $k = N(N-1)$. Practically however, the symbolic matrix power in the Neumann series gets relatively complicated for $k \sim 10$, and we usually stop the expansion at $k_{max}$ around 7 or 8. 



Similar to the $\vec{\rho}$ vector that defines the state, $\density$, operators corresponding to observables of interest can be mapped to vectors. Then, the scalar product of the vector of an observable and the vector of a state becomes the expectation value of this observable. Subsequently, a Neumann expansion for an expectation value can be inherited from that of the state vector in \Eq{\ref{eq:linear_system_soln}}. For example, if the radiated power operator $\hat{P}$ in \Eq{\ref{eq:power_output}} can be recast as a vector $\vec{V}$, then the Laplace transform of $P(t)$ is $P(s) = \vec{V} . \vec{\Rho}(s)$. The radiated power, $P(t)$, is then the inverse transformation back to the time domain. After solving the small $N$ cases, we can extrapolate the $\beta^k$ coefficient functions to a general $N$, resulting in expressions like in \Eq{\ref{eq:analytical_power}}.


\section{Symmetric system}\label{sec:continuum_approximation_sym}


In this section, we study the mean-field prediction for a simpler symmetric system to show that a MF-4 or higher is required to capture the buildup of second order coherence discussed in Sec.~\ref{sec:coherence_full_inverted}. This system has the same equations for a chiral system as in \Eq{\ref{eq:cascaded_superoperators}} but with the cascade Hamiltonian dropped; this results in a permutationally symmetric system with much simpler dynamics. This occurs when the atoms are coupled symmetrically to both the left and right propagating photons and are positioned in the mirror configuration \cite{anaPhysRevLett.131.033605}. There are other benefits of studying the permutationally symmetric system. First, it can serve as a base point to which the more complicated chiral system can be compared. These systems differ only in the coherent Hamiltonian, and thus, have the same initial dynamics \cite{robicheaux2021theoretical,anaPhysRevLett.131.033605}, which can be used as a sanity check to validate the derived equations of motion. Second, this comparison illuminates the role of chirality or a broken symmetry on the complexity and the physics of the collective light-atom interaction.
For a finite system, the mean field equations can be easily obtained via the package in Ref.~\cite{juliaplankensteiner2022quantumcumulants}. The number of equations in this case grows with the order of MF-$n$ as $\sim ~ n^3$ independently of the atom number $N$ due to the permutational symmetry of the atoms. For the case of starting from inversion with no drive, the number of equations is smaller, $\sim ~ n^2$. 

\begin{figure}[htbp]
    \centering
    \includegraphics[width=0.45\textwidth]{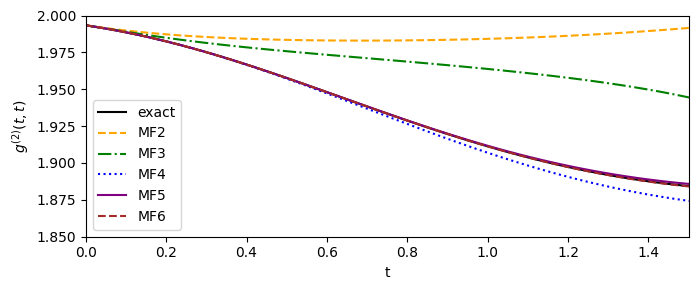}
    \caption{\justifying Comparison of the dynamics of $g^{(2)}(t,t)$ starting from the totally inverted state for a symmetric system via various orders of MF approximations and exact solution.}
    \label{fig:mf_comparison_g2tt_sym}
\end{figure}

We simulated up to MF-6 to show that a MF-4 or higher is required to capture the buildup of second order coherence. This is illustrated in Fig.~\ref{fig:mf_comparison_g2tt_sym} where MF2 and MF3 deviate appreciably from the exact solution, while MF4 or higher track the exact solution with a good agreement. This was done for the same parameters in the main text ($N = 300,\, \beta = 0.01$). Despite the difference between the symmetric and chiral systems, we note the similarity of the MF2 and MF3 predictions in this symmetric system to the symmetry-lacking chiral system as in Fig.~\ref{fig:mf_comparison_g2tt}. Such similarity of behavior between the two systems was also noted in Ref.~\cite{bach2024emergence}. Computationally, the exact solution is based on the `pisolve' method found in Ref.~\cite{permsymshammah2018open}, while the MF equations were derived and solved using the software from \cite{juliaplankensteiner2022quantumcumulants}.

\bibliography{main}

\end{document}